\documentstyle[prl,twocolumn,aps,epsf,amssymb]{revtex}

\def\bea{\begin{eqnarray}}
\def\eea{\end{eqnarray}}
\def\be{\begin{equation}}
\def\ee{\end{equation}}

\begin{document}

\title{Relaxation properties in a lattice gas model with asymmetrical particles}

\author{Anastasio D\'\i az-S\'anchez$^{1,2}$,
Antonio de Candia$^1$, and Antonio Coniglio$^1$}
\address{
$^1$Dipartamento di Scienze Fisiche, Universit\'a di Napoli
``Federico II'', \\ Complesso Universitario di Monte Sant'Angelo,
Via Cintia , I-80126 Napoli, Italy\\ and
INFM, Unit\'a di Napoli, Napoli, Italy.\\
$^2$Departamento de F\'\i sica Aplicada,
Universidad Polit\'ecnica de Cartagena, \\ 
Campus Muralla del Mar, Cartagena, 
E-30202 Murcia, Spain.}

%\date{\today}
\wideabs{%
\maketitle
\begin{abstract}
We study the relaxation process in a two-dimensional lattice gas
model, where the interactions come from the excluded volume.
In this model particles have three arms with an asymmetrical shape,
which results in geometrical frustration that inhibits full packing.
A dynamical crossover is found at the arm percolation of the particles,
from a dynamical behavior characterized by a single step relaxation above
the transition, to a two-step decay below it.
Relaxation functions of the self-part of density fluctuations
are well fitted by a stretched exponential form,
with a $\beta$ exponent decreasing when the temperature is lowered
until the percolation transition is reached, and constant below it.
The structural arrest of the model seems to happen only at the maximum
density of the model, where both the inverse diffusivity and the
relaxation time of density fluctuations diverge with a power law.
%We define a self-overlap parameter 
%which takes into account the orientation of the particles, and find 
%a two-step decay in its relaxation functions below the arm percolation.
The dynamical non linear susceptibility, defined as the fluctuations
of the self-overlap autocorrelation, exhibits a peak at
some characteristic time, which seems to diverge at the maximum density
as well.
\end{abstract}
}

\section{Introduction}

Most glassy systems such as structural glasses, ionic conductor, 
supercooled liquids, polymer, colloid, and spin glasses \cite{EJ1}
have similar complex dynamical behavior. As the temperature is lowered
the relaxation times increase drastically, and the relaxation
functions deviate strongly from a single exponential function at some 
temperature $T^*$ well above the dynamical transition. In the long time 
regime they can be well fitted by a stretched exponential 
or Kohlrausch-Williams-Watts \cite{KWW54} function
$f(t)=f_0\exp\left[-(t/\tau)^\beta\right]$, with $0<\beta\le 1$. 

There are two mechanism driving to non exponential 
relaxation. In disordered model like spin glasses, it is caused by the
existence of unfrustrated ferromagnetic-type clusters of
interactions \cite{RS85}, and therefore is a direct consequence of
the quenched disorder \cite{FC97}.
Another mechanism in frustrated systems is based on the percolation
transition of the Kasteleyn-Fortuin and Coniglio-Klein cluster \cite{FK72}.
In this approach disorder is not needed to obtain 
nonexponential relaxation \cite{FF99}.

In spin glasses there is a thermodynamic 
transition at a defined temperature $T_{SG}$, where the non linear 
susceptibility diverges. As $T_{SG}$ is approached, the static correlation 
length becomes larger. In glass forming liquids it seems that
there is no sharp 
thermodynamic transition, and no diverging static length.
However numerical studies have identified long lived dynamical 
structures which are characterized by a typical length and a typical 
relaxation time, which depend on temperature and density \cite{DH98}. In order 
to characterize this behavior a dynamical nonlinear susceptibility was
introduced by Donati et al. \cite{DF99}, both for spin models and for 
structural glasses. They have found that the dynamical susceptibility 
exhibits a maximum at some characteristic time which diverges as the
dynamical transition temperature is approached from above. Also in the
annealed version of the frustrated Ising lattice gas model \cite{C94}
a similar behavior has  been found, while in the quenched
version the dynamical susceptibility is always increasing, due to
the presence of the thermodynamic transition \cite{FC00}.

The understanding of the macroscopic process of relaxation in structural
glasses, starting from the microscopic motion of the particles,
has been boarded using different microscopic models, as for example
the hard square model \cite{GF66}, the 
kinetically constrained models \cite{FA84,GP97}, and the frustrated Ising 
lattice gas (FILG) model \cite{C94}.
These models have reproduced some aspects of the glassy 
phenomenology, and recently the FILG model has been 
studied in its quenched and annealed version with kinetic constraints
\cite{FC00}. The results found in 
the annealed version seems to be closer to the experimental ones.
In this paper we consider a two-dimensional grometrical model,
which contains as main ingredients only geometrical frustration without
quenched disorder and without kinetic constraints, as quenched disorder
is not appropriate to study structural glasses and kinetic constraints 
are some how artificial. Similar models have already been proposed \cite{C94}
and applied to study granular material \cite{CL97}.

In Sec. II we present the model and in
Sec. III we study its percolation properties. In Sec. IV we show the
dynamical results, and finally in Sec. V we present our conclusions.

\section{The model}

In this paper we introduce a model which can be considered as an 
illustration of the concept of frustration arising as a packing problem. 
In systems without underlying crystalline order, frustration is typically
generated by the geometrical shape of the molecules, which prevents the 
formation of close-packed configurations at low temperature or high density; 
for systems with underlying crystalline order, frustration arises when the 
local arrangement of molecules kinetically prevents all the molecules from 
reaching the crystalline state.

An example of glass-former that has difficulty in achieving crystalline order
is the ortho-terphenyl, whose molecule is made of
three rings. This system can be loosely modeled with a simple lattice
model, in which ``T'' shaped objects occupy the vertices of a square
lattice with one of four possible orientations.

Assuming that the arms cannot 
overlap due to excluded volume, we see that only for some relative
orientations 
two particles can occupy nearest-neighbor vertices.
Consequently, depending on the local arrangement of particles,
there are sites on the lattice that cannot be occupied (see Fig. \ref{fig:1}).
This type of ``packing'' frustration thus induces defects or 
holes in the system. This model resembles the hard-square lattice gas 
model \cite{GF66} which can be seen as ``+'' shaped objects on the 
vertices of a square lattice with excluded volume interaction. A very 
important difference between these two models is the internal degree of
freedom due to the particle shape which is absent in the latter.

We consider a two dimensional square lattice and impose periodic boundary 
conditions. In our system the maximum of density is $\rho_{\rm max}=2/3$ 
at which all possible bonds are occupied by an arm. 
A configuration of density $\rho_{\rm max}$ is a ground state of the
system, corresponding to chemical potential $\mu\rightarrow\infty$
or temperature $T\rightarrow 0$. It can be obtained 
for any size constructing larger systems from smaller 
ones with $\rho_{\rm max}$ and appropriated boundary conditions.
In this way one can build an extensive number of different ground states
that lack spatial order.

We have simulated the diffusion and rotation dynamics of this model by Monte 
Carlo methods. The dynamics of the particles is given by the
following algorithm:
i) Pick up a particle at random;
ii) Pick up a site at random between the four nearest neighbor ones;
iii) Choose randomly an orientation of the particle;
iv) If it does not cause the overlapping of two arms,
move the particle in the given site with the given orientation;
v) If the diffusion movement is not possible,
choose a random orientation and try to rotate the particle to this new
orientation;
vii) Advance the clock by $1/N_s$, where $N_s$ is the number 
of sites, and go to i).

Studying the dynamics by Monte Carlo simulations, the finite size effects 
are larger when we are near to $\rho_{\rm max}$, because the 
particles can be enclosed in cages, and the diffusion is blocked. In 
lattices of finite size, cages may be indefinitely stable. This has also 
been observed in the hard-square lattice gas model \cite{GF66} where at 
the maximum of density the particles are on the diagonals, but in lattices 
of finite size cages of particles indefinitely stable are formed at lower 
densities. 

\section{Percolation transition}
 
In order to investigate whether the percolation transition has
effects on the dynamics, 
in this section we analyze the percolation properties of the model,
and relate the percolation density with a change in the dynamical 
properties of the model.
The particles have three arms, and there are two bonds per site on the
lattice, so the density of bonds occupied by an arm is given by
$\sigma =3\rho/2$, where $\rho$ is the density of particles.
Therefore, if the correlations of the arms were not important,
the arm percolation would occur at the density $\sigma_c=1/2$,
corresponding to $\rho_c=1/3$. Nevertheless we expect some correlation
effects.

We have simulated our system for various lattice sizes around the 
percolation density in order to determine $\rho_c$. For each density we 
have reached equilibrium and then, taking $10^4$ steps, we have evaluated 
the probability of existence of a spanning cluster $P$ and the mean cluster 
size $S=\sum_s s^2 n_s$,
where $n_s$ is the density of finite clusters of size $s$.

Around the percolation density the averaged quantities $P(\rho)$ and 
$S(\rho)$, for different values of the lattice size $L$, should obey to the 
finite size scaling \cite{SA94}
\bea
P(\rho)&=&F_P\left[L^{1/\nu}(\rho-\rho_c)\right]\\
S(\rho)&=&L^{\gamma/\nu}F_S\left[L^{1/\nu}(\rho-\rho_c)\right]
\eea
where $\gamma$ and $\nu$ are then critical exponents,
and $F_P(x)$ and $F_S(x)$ 
are universal functions of an adimensional quantity $x$. Fig. \ref{fig:2}
shows 
the finite size scaling of $P$ and $S$. We have selected the values of the 
exponent corresponding to the two-dimensional site-bond percolation 
universality class \cite{SA94}, that is $\nu=4/3$ and $\gamma=43/18$ and
we have found $\rho_c=0.315 \pm 0.003$.
The critical density can also be determined from $P(\rho)$ as 
the density at which curves corresponding to different sizes cross.
From the inset of Fig. \ref{fig:2}(a) we see that $\rho_c$
is between 0.315 and 0.32, so that the 
arm correlations and the thermal process decrease the critical density
with respect to the random bond problem.

\section{Dynamical results}

We wish now to define a self-overlap parameter, in order to measure
the autocorrelation of the model.
Different definitions of $q$ can be used, and
the form of the relaxation functions is different for different overlap 
parameters. Here we define a self-overlap parameter 
which takes into account the orientation
of the particle, rather that its position.
The orientation of the particle is defined 
by the discrete values of the angle, $\phi_i=0$, $\pi/2$, $\pi$,
or $3\pi/2$, and we define the self-overlap
\be
q(t)=\frac{1}{N}\sum_i n_i(t')n_i(t'+t)
\cos\left[\phi_i(t'+t)-\phi_i(t')\right] 
\ee
where $n_i(t)=0,1$ is the occupation number of site $i$ at time $t$,
$\phi_i(t)$ is the orientation of the particle on site $i$
at time $t$, and $N$ is the number of sites. This parameter is a 
generalization of the self-overlap defined in Ref. \cite{FC00},
where the orientation  $\phi_i$
plays the role of the spin variables.
When all particles are frozen then $q=1$. 
The associated dynamical nonlinear susceptibility $\chi(t)$ is given by
\be
\chi(t)= N \left[\langle q(t)^2\rangle -\langle q(t)\rangle ^2\right],
\ee
where the average $\langle\cdots\rangle$ is done on the reference time $t'$.
 
In Fig. \ref{fig:3}
we show the relaxation function of the self-overlap parameter, for
a system of size $64^2$ and densities between $\rho=0.2$ and $0.66$. Each 
curve is obtained averaging over a time interval for $t'$ of $10^4-10^8$ 
Monte Carlo steps, depending of the density. For high and intermediate 
densities we observe a two step relaxation functions, which gives 
evidence of two well separate time scales in the system. The first short 
time decay of the relaxation functions is due to the rotations of the 
particles in a frozen environment, which appears as quenched on this time 
scale, while the second decay is due to the evolution of the environment 
and final relaxation to equilibrium.

For very low densities $\rho\le 0.1$
the relaxation function is well fitted
by an exponential form. From $\rho \approx 0.1$ to
$\rho \approx 0.32$ we can fit
the relaxation function by a stretched exponential form, with the exponent
$\beta$ decreasing slowly until $\beta\approx 0.9$ at $\rho\approx 0.32$. 
For higher values of densities it is impossible to fit the
relaxation function with a single form over the whole time interval.
As we have seen, $\rho\approx 0.32$ corresponds to arm percolation.
This suggests that the percolation transition induces a crossover 
in the dynamics, from a single relaxation regime to two temporal scales
for the relaxation of the particles.
At densities higher than $\rho\approx 0.32$, the long time tail of the 
relaxation functions is well fitted by a stretched exponential form, where 
the exponent $\beta$ depends very weakly from the density (it is constant 
within the errors) and ranges between $\beta=0.64$ and $\beta=0.71$. We 
show in Fig. \ref{fig:4}
the time-density superposition of the relaxation function
$\langle q(t) \rangle$, for densities between $\rho=0.62$ and $\rho=0.66$.
Also shown as a dashed line is a fit with a stretched exponential function.

In Fig. \ref{fig:5}
we show the intermediate time behavior of the self-overlap relaxation 
functions, for densities $\rho=0.62$, $0.64$ and $0.65$.
We have tried to fit them with the 
simplified form of the function predicted by the MCT,
\begin{equation}
\langle q(t) \rangle = f+At^{-a}-Bt^b
\end{equation}
where the fitting parameters are $f$, $A$, $B$ and $\lambda$, while $a$
and $b$ are related to $\lambda$ by the transcendental equation
\be
\frac{\Gamma^2(1-a)}{\Gamma(1-2a)}=
\frac{\Gamma^2(1+b)}{\Gamma(1+2b)}=\lambda.
\label{eq:trash}
\ee
The values of $\lambda$ given from the fits are constant within the errors,
with a mean value $\lambda=0.785 \pm 0.005$, which correspond to
exponents $a=0.285 \pm 0.005$ and $b=0.50 \pm 0.01$.

In Fig. \ref{fig:6}
we show the dynamical non linear susceptibility for some values 
of the density. The maximum in the susceptibility $\chi(t^*)$ and the time of 
the maximum $t^*$ seem to diverge together when the density grows. This has 
also been found previously in other models as p-spin, Lennard-Jones binary 
mixture \cite{DF99}, and in the annealed version of the frustrated Ising 
lattice gas model \cite{FC00}. In our model we obtain that the maximum 
of $\chi(t^*)$ can be fitted by the power law 
$\chi(t^*) \propto (\rho_{\rm max}-\rho)^{-\alpha}$. Here we have 
$\rho_{\rm max}=0.664\pm 0.002$ and $\alpha=0.71\pm 0.02$. The equilibrium
value is $\chi(t \rightarrow \infty)=\rho^2/2$ for low densities and
$\chi(t \rightarrow \infty)=1/2$ for the higher ones.

The density-density autocorrelation function and its dependence with the 
time is an important property which characterizes the glassy behavior. We have 
studied the self-part of the autocorrelation function of the density  
fluctuations defined as
\be
F_{\bf k }^s(t)=\frac1N\left< \sum_i 
e^{i{\bf k }({\bf r}_i(t'+t)-{\bf r}_i(t'))}\right>,
\ee
where ${\bf r}_i(t)$ is the position of the $i$th particle in units of 
the lattice constant. The wave vector can take the values 
${\bf k }=(2\pi/L){\bf n }$, where ${\bf n }$ has integer components $n_x$ and 
$n_y$ ranging from 0 to $L/2$.
 
Fig. \ref{fig:7}
shows $F_{\bf k }^s(t)$ corresponding to $k_x=\pi$ and $k_y=0$ for 
different densities. For all densities the whole time interval of the 
autocorrelation function can be fitted by a stretched exponential function, 
$f(t)=\exp\left[-(t/\tau)^\beta\right]$, where the
exponent $\beta$ depends of the 
density. In Fig. \ref{fig:8} we show $\beta$ as a
function of the density. We can see that the 
exponent $\beta$ decreases with the density until a density near 
$\rho \approx 0.32$ is reached. From this density the exponent 
becomes constant $\beta \approx 0.82$ (within of the error bars). 
At densities near to $\rho_{\rm max}$ 
the finite size effects become important and $\beta$
deviates from the constant value.
As in the $\alpha$-relaxation of the self-overlap parameter, 
the effect of the percolation on the relaxation of $F_{\bf k }^s(t)$ is to 
leave $\beta$ constant, but now there is not two step relaxation process.

If we look at the relaxation of the position of particles there is no
vibration relaxation, the relaxation begins
when frozen domains evolve. Instead the relaxation of the orientation
of the particles has a relaxation process where it is possible
the vibration of the rotational degree of freedom. Nevertheless after
the arm percolation the position and orientation relaxation in the
$\alpha$-relaxation have an stretched exponential form with the exponent
$\beta$ constant (within the errors) although with different value 
each other. 

The relaxation time $\tau$ is obtained from the
fitting of $F_{\bf k }^s(t)$ with
a stretched exponential function, and  can be fitted by
a power law $\tau\propto (\rho_{\rm max}-\rho)^{-\gamma}$, with 
$\gamma=2.70 \pm 0.02$ and $\rho_{\rm max}=0.666 \pm 0.003$
(see Fig. \ref{fig:9}).
Note that the value of $\gamma$ coincides within the errors with the
one deduced by the exponents $a$ and $b$ given by the fit of Fig.
\ref{fig:5}, using the relation predicted by Mode-Coupling Theory
\be
\gamma=\frac{1}{2a}+\frac{1}{2b},
\ee
which gives $\gamma=2.74\pm0.03$.

We have calculated the diffusion coefficient from the 
mean-square displacement $\langle\Delta r(t)^2\rangle$ at very long times. 
The values obtained for $D$ are well fitted by a power law near of 
$\rho_{\rm max}$, $D\propto (\rho_{\rm max}-\rho)^\gamma$
with $\gamma=2.70 \pm 0.02$ and $\rho_{\rm max}= 0.666 \pm 0.003$ (see 
the inset of Fig. \ref{fig:9}).
The power law agrees with the law found in some lattice gas 
models (the kinetically constrained model \cite{FA84}) but it is different 
of the law found in the hard-square lattice gas model \cite{GF66}.
This singular 
behavior of $D$ is in accordance with the prediction of MCT.
Form the behavior of $D$ and $\tau$ we arrive to
$D^{-1} \propto \tau$ and so to the Stokes-Einstein relation
$D^{-1}\propto \eta$ (due to $ \tau \propto \eta$).

\section{Conclusions}

We have proposed a two-dimensional geometrical mo\-del,
based on the concept of geometrical
frustration which is generated by the particle shape.
This model 
has neither quenched disorder nor kinetic constraints.
Percolation in the model has been studied, and it has been found that
at the critical density of arm percolation the model shows a 
dynamical crossover, characterized by the onset of a two step relaxation
in the orientational relaxation of the particles.
Below the transition, the long time regime of orientational and 
positional relaxation functions can be fitted by a stretched exponential
with an exponent $\beta$ constant within the errors.
The dynamical results have been compared with the
prediction of the MCT, finding good agreement.
A self-overlap parameter has been
defined which takes into account the orientation of particles,
and the corresponding dynamical non linear susceptibility has been studied.
It has been found that this dynamical susceptibility shows a peak at some 
characteristic time, which gives evidence of long lived dynamical
structures with a growing length and relaxation time, as found in some
molecular dynamics simulations.

\acknowledgments

This work was supported in part by the European TMR Network-Fractals
(Contract No. FMRXCT\-980183), MURST-PRIN-2000 and INFMPRA(HOP).
We acknowledge the
allocation of computer resources from INFM Progetto Calcolo Parallelo.
A. D\'\i az-S\'anchez acknowledges support
from a Postdoctoral Grant from the European
TMR Network-Fractals.

\newpage

\def\mysize{5cm}

\begin{figure}
\epsfysize=\mysize
\begin{center}
\epsfbox{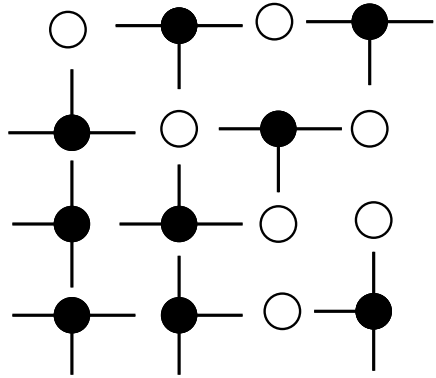}
\end{center}
\caption{Schematic picture of one particular configuration in a system 
size of $4^2$ and density $\rho=9/16$.} 
\label{fig:1}
\end{figure}

\begin{figure}
\epsfysize=\mysize
\begin{center}
\epsfbox{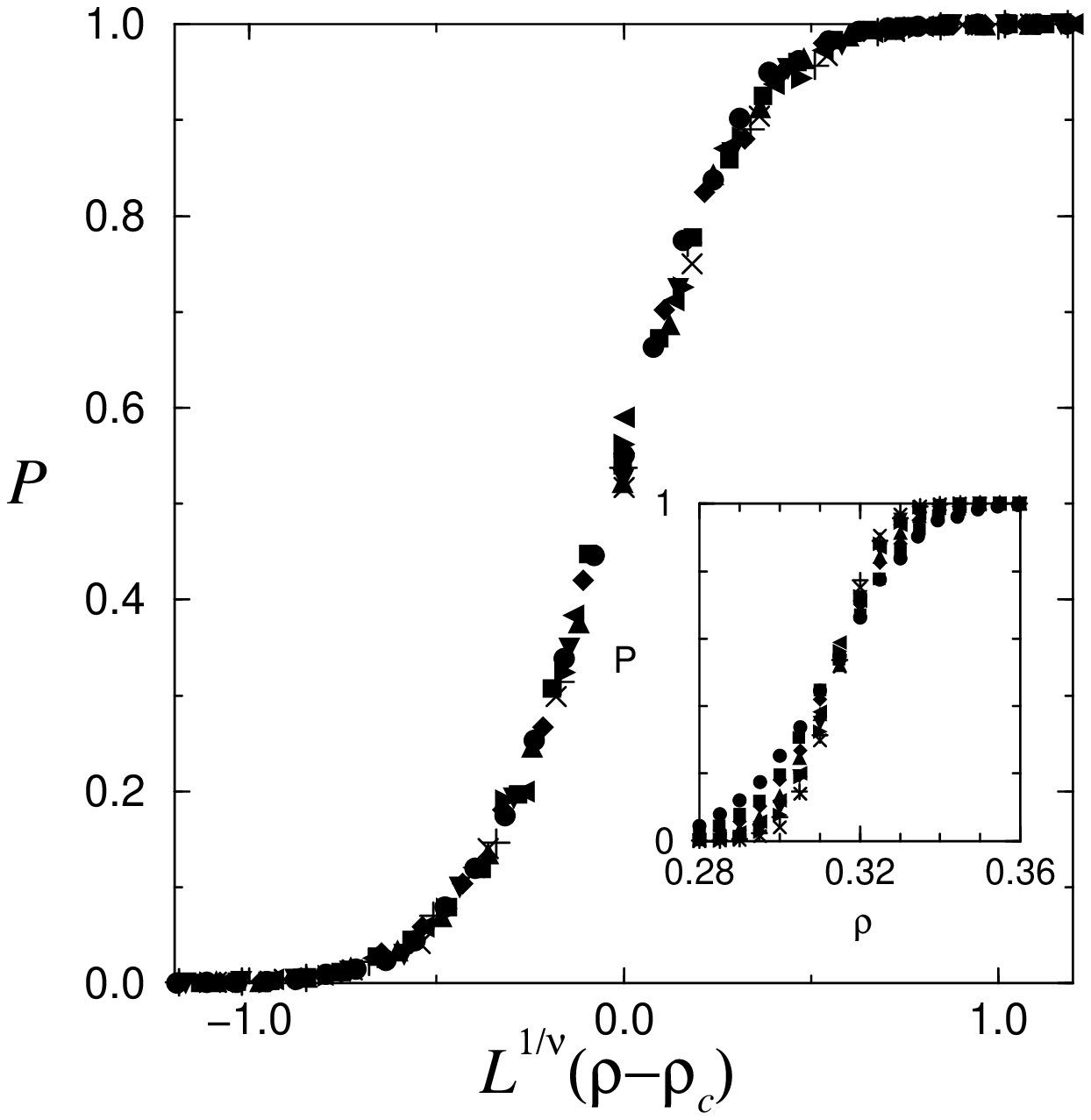}
\end{center}
\epsfysize=\mysize
\begin{center}
\epsfbox{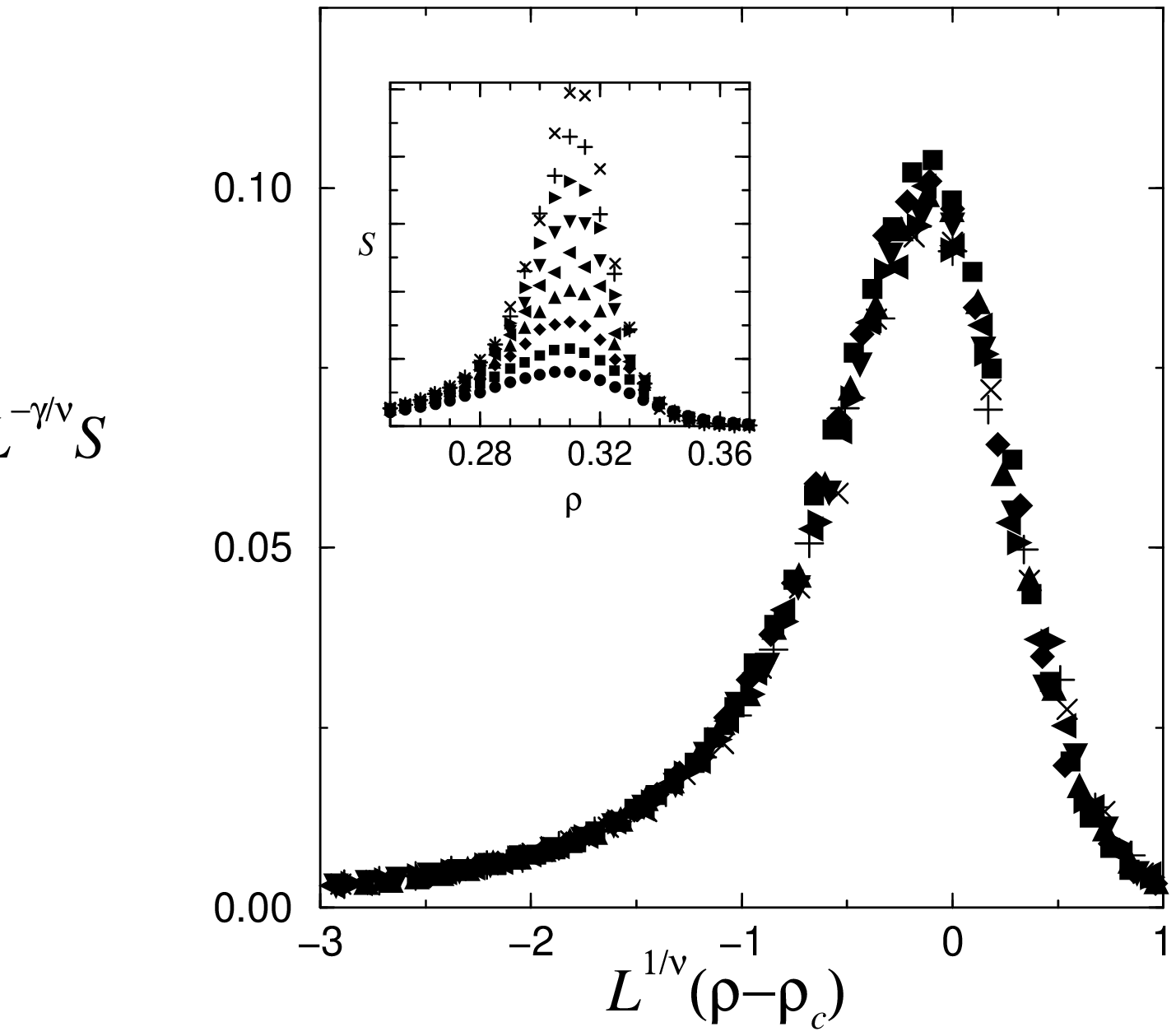}
\end{center}
\caption{Finite size scaling of (a) $P(\rho)$ and
(b) $S(\rho)$ for lattice sizes
$L=40$, 50, 60, 70, 80, 90, 100, 110, and 120. Inset: (a) $P(\rho)$ and 
(b) $S(\rho)$ for the same sizes.} 
\label{fig:2}
\end{figure} 

\begin{figure}
\epsfysize=\mysize
\begin{center}
\epsfbox{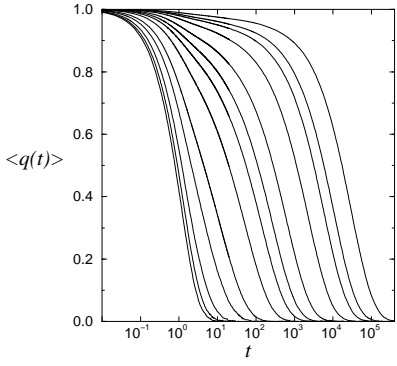}
\end{center}
\caption{Relaxation functions for the
self-overlap for system size $64^2$ and densities
$\rho=0.2$, 0.3, 0.4, 0.5, 0.55, 0.6,
0.62, 0.63, 0.64, 0.65, 0.655, 0.657, and 0.66.} 
\label{fig:3}
\end{figure}      

\begin{figure}
\epsfysize=\mysize
\begin{center}
\epsfbox{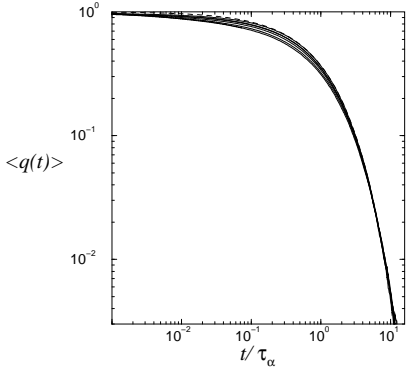}
\end{center}
\caption{Time-density superposition principle
for the relaxation function of the 
self-overlap, for densities $\rho=0.62$, 0.63,
0.64, 0.65, 0.655, 0.657, and 0.66. The dashed
line is a fitting function corresponding to a
stretched exponential form with exponent 
$\beta=0.71$.} 
\label{fig:4}
\end{figure}      

\begin{figure}
\epsfysize=\mysize
\begin{center}
\epsfbox{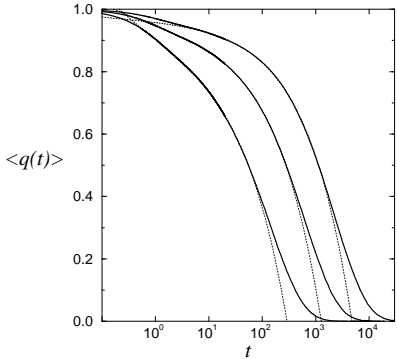}
\end{center}
\caption{Fit of the intermediate time regime of the relaxation functions of 
the self-overlap, for densities $\rho=0.62$, $0.64$ and $0.65$.
The fitting function is $f+At^{-a}-Bt^b$ (dotted line), where the fitting
parameters are $f$, $A$, $B$, and $\lambda$, and $a$ and $b$ are
given by the relation (6).}
\label{fig:5}
\end{figure}

\begin{figure}
\epsfysize=\mysize
\begin{center}
\epsfbox{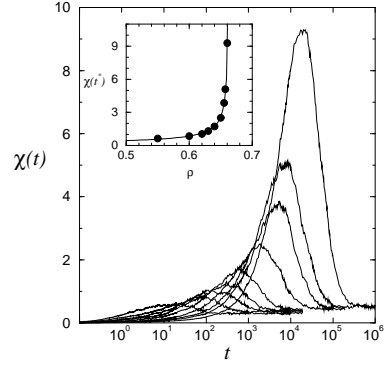}
\end{center}
\caption{Dynamical susceptibility for $L=64$ and densities $\rho=0.6$, 0.62, 
0.63, 0.64, 0.65, 0.655, 0.657, and 0.66. Inset: the maximum $\chi(t^*)$ as a
function of density. The fitting function is a power law 
$\chi(t^*)=0.12(0.664-\rho)^{-0.7}$.} 
\label{fig:6}
\end{figure}

\begin{figure}
\epsfysize=\mysize
\begin{center}
\epsfbox{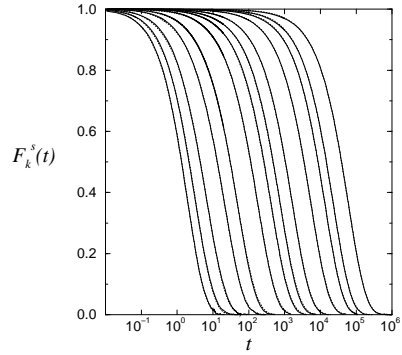}
\end{center}
\caption{Relaxation functions for the self-part of the density-density 
autocorrelation function for $k_x=\pi$ and $k_y=0$ and the same system size
and densities of Fig. 2. Dotted lines are fitting functions corresponding to
stretched exponential functions.} 
\label{fig:7}
\end{figure} 

\newpage

\begin{figure}
\epsfysize=\mysize
\begin{center}
\epsfbox{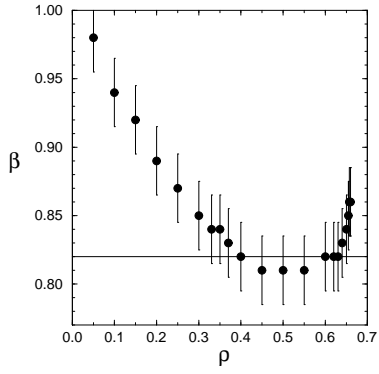}
\end{center}
\caption{Parameter $\beta$ as a function of the density, obtained by fitting 
$F_{\bf k }^s(t)$ for $k_x=\pi$ and $k_y=0$ with the function 
$f(t)=\exp\left[-(t/\tau)^\beta\right]$. The densities represented are 
$\rho=0.05$, 0.1, 0.15, 0.2, 0.25, 0.3, 0.33, 0.35, 0.37, 0.4, 0.45, 0.5, 0.55, 
0.6, 0.63, 0.64, 0.65, 0.655, and 0.66.
The solid line corresponds to $\beta=0.82$.} 
\label{fig:8}
\end{figure} 

\begin{figure}
\epsfysize=\mysize
\begin{center}
\epsfbox{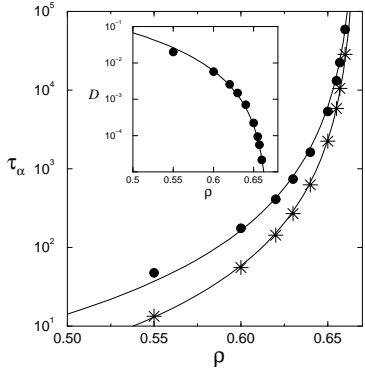}
\end{center}
\caption{The relaxation time $\tau_\alpha$ as a function of density $\rho$.
Solid circles: relaxation times obtained from the fit in Fig. \ref{fig:7}.
Asterisks: relaxation times obtained from the fit in Fig. \ref{fig:4}.
The
solid lines are power law functions $\tau_\alpha=a(0.666-\rho)^{-2.7}$,
with $a=0.115$ in the former case and $a=0.04$ in the latter.
Inset: the diffusion constant $D$ as function of density $\rho$. The fitting 
function is a power law $D=8(0.666-\rho)^{2.7}$.} 
\label{fig:9}
\end{figure} 

\end{document}